\documentclass{aa}
\usepackage{times}
\usepackage{epsfig}
\usepackage{colordvi}

\begin{document}

   \thesaurus{11    
              (11.01.2; 
	       11.05.1; 
               11.17.1; 
               11.17.4; Q2112+059)}
\title{Deep imaging of Q2112+059:
A bright host galaxy but no DLA absorber
             \thanks{
Based on observations made with the Nordic Optical Telescope,
operated on the island of La Palma jointly by Denmark, Finland,
Iceland, Norway, and Sweden, in the Spanish Observatorio del
Roque de los Muchachos of the Instituto de Astrofisica de
Canarias.}
\fnmsep\thanks{Based on observations with the NASA/ESA Hubble Space
Telescope, obtained from the data Archive at the Space Telescope 
Science Institute,
which is operated by the Association of Universities for Research
in Astronomy, Inc. under NASA contract No.  NAS5-26555.}
}

\author{J.U. Fynbo
        \inst{1}
        \and
        P. M\o ller
        \inst{1}
        \and
        B. Thomsen
        \inst{2}
        }

\offprints{J.U. Fynbo}
\mail{jfynbo@eso.org}

\institute{
           European Southern Observatory, Karl-Schwarzschild-Stra\ss e 2,
           D-85748, Garching by M\"unchen, Germany
           \and
           Institute of Physics and Astronomy,
           \AA rhus University, DK-8000 \AA rhus C
           }

\date{Received ; accepted }

\maketitle

\begin{abstract}
  In a ongoing programme aimed at studying galaxy counterparts of Damped
Ly$\alpha$ Absorbers (DLAs) we have obtained high resolution deep I-band
imaging data of the field around the z$_{em}$ = 0.457 BAL QSO
Q2112+059. In the literature this QSO is listed to have a candidate DLA at 
z$_{abs}$ = 0.2039 along the line of sight. After subtraction of the QSO 
Point Spread Function (PSF) we detect a galaxy centred on the position of 
Q2112+059. To help answer whether this galaxy is the DLA or the QSO host 
galaxy we retrieved a GHRS spectrum of Q2112+059 from the HST-archive. This 
spectrum shows that there is no Ly$\alpha$ absorption line at z$_{abs}$ = 
0.2039. This fact in combination with the perfect alignment on the 
sky of the galaxy and Q2112+059 lead us to the conclusion that the galaxy 
must be the host galaxy of Q2112+059.

The host galaxy of Q2112+059 is bright (M$_I^{obs} = -23.6$), and
has a radial profile well fitted by a {\it modified Hubble + de
Vaucouleurs} profile with $R_c = 0.5$kpc and $R_e = 3.6$kpc.
Our results are well in line with the conclusion of earlier
work done at lower redshifts, that bright low redshift QSOs
preferentially reside in luminous, elliptical galaxies.
The host of Q2112+059 is however, despite it's brightness,
very compact when compared to early type galaxies at lower redshifts.

\end{abstract}

%

\section{Introduction}

We have during the last several years been undertaking programmes
aimed at detecting optical counterparts of the most \ion{H}{I} rich 
class of QSO absorption line systems, the Damped Ly$\alpha$ absorbers 
or DLAs for short
(M\o ller and Warren 1993, 1998; Fynbo et al. 1999, 2000a,b).
DLAs cause Ly$\alpha$ absorption lines with damping wings corresponding
to \ion{H}{I} column densities larger than $2\times10^{20} cm^{-2}$. This very
large column density absorption occurs in regions of self shielding,
cool gas, i.e. where one would expect stars to form. Furthermore,
they contain nearly all the neutral gas in the universe (Wolfe et al.
1995). Hence DLAs at high redshift are prime candidates for being the
progenitors of present day galaxies.

At redshifts $z<1.65$ DLAs are both rarer and observationally more difficult
to detect since Ly$\alpha$ is not shifted above atmospheric cut--off
and hence cannot be observed with ground based telescopes
(Lanzetta et al. 1995; Jannuzi et al. 1998; Rao \& Turnshek 2000).
Le Brun et al. (1997) reported candidate galaxy counterparts
of 7 such low redshift absorbers that are either confirmed DLAs
or strong MgII/FeII absorbers. They found that those candidate galaxy
counterparts cover a wide range in morphological types.

 The z$_{em}$ = 0.457 BAL QSO \object{Q2112+059} (2000.0 coordinates 
RA 21 14 52.6, DEC +06 07 41) is radio quiet, and
has optical magnitudes B = 15.52 and V = 15.77. Lanzetta et al. (1995) 
reported the detection of a candidate DLA at z$_{abs}$ = 0.2039 in the 
IUE spectrum of Q2112+059. In September 1996 we performed an imaging 
study of Q2112+059 in an attempt to detect the optical counterpart 
of this DLA. The imaging and data reduction is described in Sect. 2, 
in Sect. 3 we present spectroscopic data from the Hubble Data
Archive and in Sect. 4 and 5 we discuss our results.

For easy comparison to earlier work on this subject, we shall here
assume a Hubble constant of 50 km s$^{-1}$ Mpc$^{-1}$ and $\Omega$=1.0.

\section{Imaging}

\subsection{Observations and Data Reduction}
The observations were performed in 1996 with HiRAC (High Resolution
Adaptive Camera) on the 2.56 m Nordic Optical Telescope under photometric
and excellent seeing conditions. The CCD used was a 1k$\times$1k back-side 
illuminated thinned Tektronix with a pixel scale of 0.176 arcsec. The data 
were all obtained in the Bessel I filter. The journal of observations is 
given in Table~\ref{Journal}. A total integration time of 3.25 hours
was obtained.
Also observed were 7 Landolt sequences (Landolt 1992).
\begin{table}[b]
\caption{Journal of NOT observations}
\begin{center}
\begin{tabular}{@{}lcccccc}
Date       & Seeing FWHM   & Number of & Exp. Time \\
           & (arcsec) & Frames    &  (sec)        \\
\hline
1996 Sep15 & 0.6-0.8  &  7        &  2100         \\
1996 Sep17 & 0.5-0.7  &  10       &  3000         \\
1996 Sep18 & 0.6-0.8  &  13       &  3900         \\
1996 Sep20 & 0.5-0.7  &  4        &  1200         \\
1996 Sep21 & 0.8-0.9  &  5        &  1500         \\
\hline
\end{tabular}
\label{Journal}
\end{center}
\end{table}

The 39 individual frames were bias subtracted, flat-fielded and
subsequently coadded using the optimal combination code described in
M\o ller and Warren (1993).

\begin{figure}[t]
 \begin{center}
 \epsfig{file=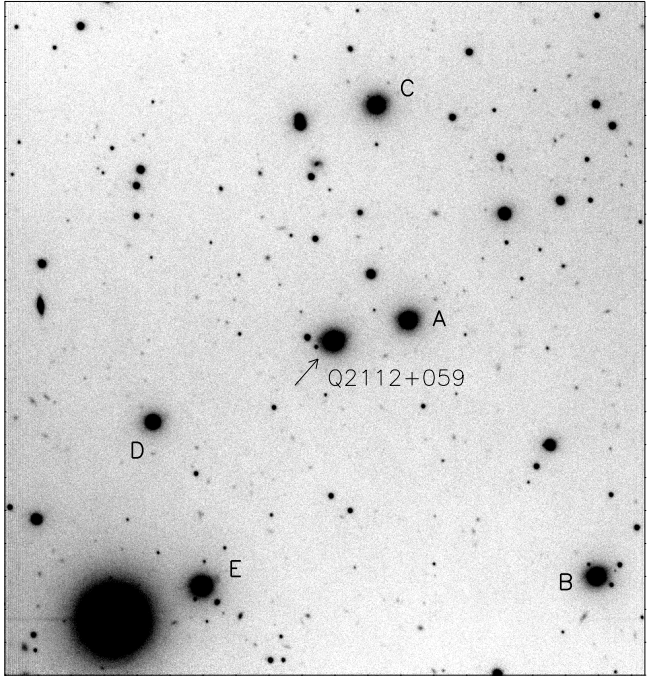,width=8.5cm}
 \end{center}
 \caption{Combined frame containing 973$\times$1023 pixels (171$\times$180
 arcsec$^{2}$) surrounding the QSO Q2112+059 (marked by an arrow). North 
 is up and east to the left. Also marked are the field stars A--E which
 we use to check the stability of the PSF in Sect. 2.2.}
\label{field}
\end{figure}

In Fig.~\ref{field} we show the final combined image of 973$\times$1023 
pixels (171$\times$180 arcsec$^{2}$) surrounding Q2112+059 which is marked 
with an arrow.
    Using the Landolt sequences we obtained the calibration from the
instrumental system to Cousins I. The uncertainty in the
zero--point is 0.02 mag. For Q2112+059 we find I = 15.16$\pm$0.02, where
the error is dominated by the zero--point error. The sky brightness was
$\mu_I$(sky) = 19.0 mag arcsec$^{-2}$ and the sky noise in the final
combined image 26.8 mag arcsec$^{-2}$. The
full--width--at--half--maximum (fwhm) of point sources in the
combined image is 0.69 arcsec.

There is no indication of variability of Q2112+059 with amplitude above
0.01mag on time-scales from tens of minutes to days.

\subsection{PSF quality and stability test}
\begin{figure}
\begin{center}
\epsfig{file=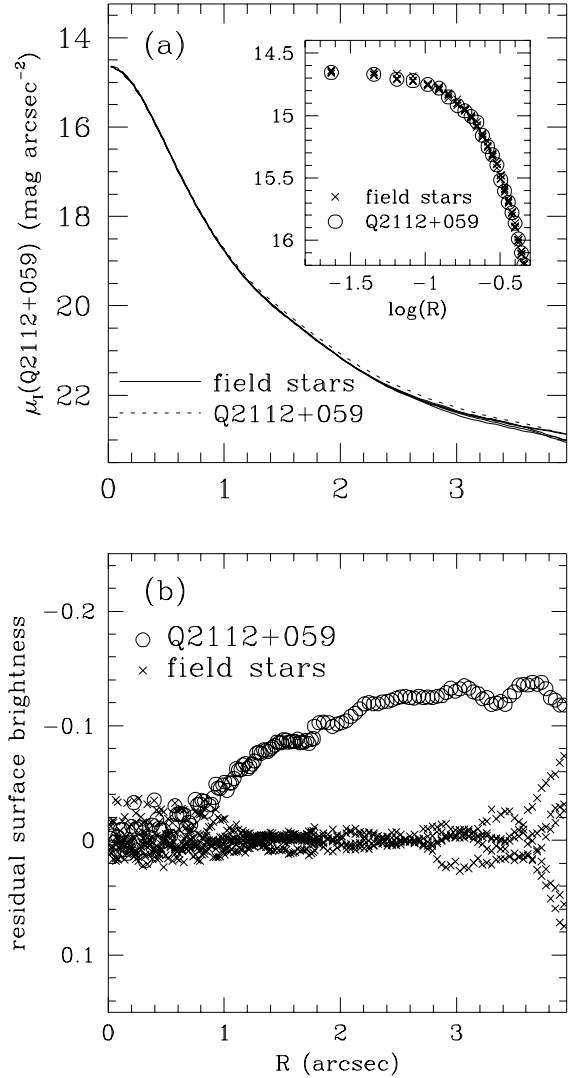,width=8.5cm}
\end{center}
\caption{-- {\bf a)} The radial profiles of 5 bright field stars and
of the QSO. The profiles were determined as the azimuthal
average on a 9 by 9 sub--pixel grid. The profiles have been normalised
to the same value in the centre (illustrated in the insert where we
plot the central part of the $\mu_{\rm I}$ profiles vs. log(R).
The QSO clearly has excess extended flux compared to the stellar
profiles.
{\bf b)} Residuals after subtraction of the weighted
mean stellar profile determined from the stars A--E. The residuals
for the 5 stars are 0 to within the noise, but
there is a clearly detected residual for Q2112+059.}
\label{PSF}
\end{figure}

In order to search for faint sources near the line of sight to 
the bright Q2112+059 a good PSF-subtraction is mandatory. 
To test the stability of the PSF we selected 5 bright stars 
distributed over the full field. These stars are 
marked A--E in Fig.~\ref{field}. We then determined the radial profiles
of the field stars by averaging over the azimuth angle on a 9 by 9
sub--pixel
grid. In Fig.~\ref{PSF}a we have plotted the radial profiles determined
in this way. Also plotted is the radial profile of Q2112+059. All
profiles were normalised to the same level in the centre (shown 
in detail in the inserted logarithmic plot in Fig.~\ref{PSF}a).
The surface brightness ($\mu_{\rm I}$) zero--point in
Fig.~\ref{PSF}a was chosen to be that of the Q2112+059 profile.
It is seen that, 
{\it i)} the 5 bright stars have radial profiles (full drawn lines) that
are identical to within the thickness of the lines out to about
2.8 arcsec, and
{\it ii)} the radial profile of Q2112+059 (short dash) shows excess
flux over that of the field stars. To quantify the excess we plot in
Fig.~\ref{PSF}b the residual surface brightness after subtraction of the
weighted mean stellar profile determined from the field stars A -- E.
At radial distances ranging from 1.4 to 3.6 arcsec from the centroid
of the QSO, the excess surface brightness is 0.08--0.14 mag.
In the same range the scatter of the stellar PSFs is about a factor
10 -- 20 smaller. At radial distances larger than 3.6 arcsec the
profiles are dominated by sky noise.

We end this section concluding that the PSF is very stable
across the field and that the profile of Q2112+059 has significant
excess flux over that of a PSF. 

\subsection{2D PSF--subtraction}
\begin{figure}[t]
\begin{center}
\epsfig{file=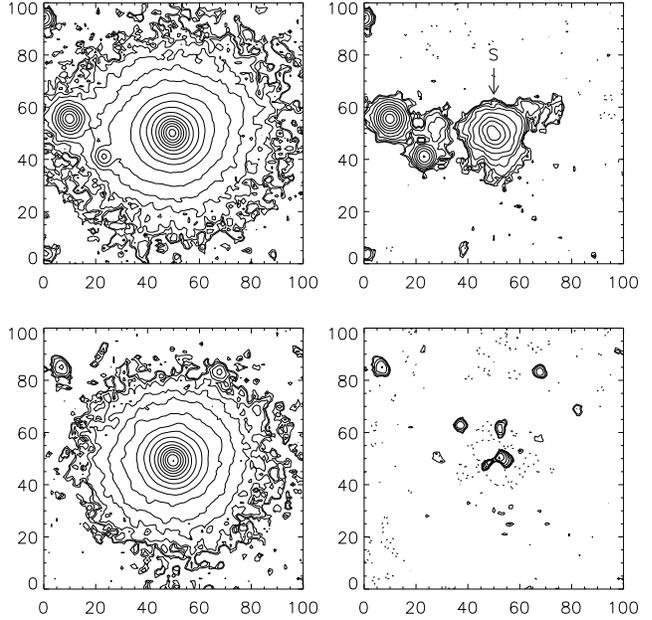,width=8.5cm}
\end{center}
\caption{{\it Upper left} : 100$\times$100 pixels
(17.6$\times$17.6 arcsec$^{2}$) surrounding
Q2112+059. {\it Upper right} : Same as left frame, but after subtraction
of the 2D PSF. {\it Lower panels} : For comparison we show the field
star marked C in Fig. 1 before and after PSF--subtraction.
The images have all been smoothed by a
3$\times$3 box car filter. The contour levels are
-4, 4, 6, 8, 12, 20, 38,... in terms of sigma of the sky in the
unfiltered image, with the negative contour being dotted.}
\label{fig3}
\end{figure}

To examine the nature of the excess flux we proceeded with a two
dimensional PSF subtraction using DAOPHOT II (Stetson 1987, 1999). For
the determination of the 2D PSF we chose to use star A, which is the
closest to Q2112+059, but as demonstrated above our results are not 
dependent on this specific choice of PSF star.
When we subtracted the PSF of the QSO using the scaling of the PSF
automatically chosen by DAOPHOT II this produced a residual
consisting of an extended positive residual surrounding a negative
``hole'' in which the PSF had been over--subtracted. This is the 
typical residual one will see when a point source and an extended 
object are superimposed on roughly the same position (see e.g. Fig.
2b where the PSF subtracted QSO profile is 0 in the centre but
0.13 mag 3 arcsec out). To obtain a image of the underlying extended
object without the ``hole'' from the PSF subtraction, we must
impose some additional condition. A reasonable assumption is that the
extended object is ``flat'' across the centroid of the QSO, so we shall use
this as the extra constraint to break the degeneracy.

Fig.~\ref{fig3} shows the result of this final two dimensional 
PSF--subtraction. In the upper left panel is shown 100$\times$100 pixels
(17.6$\times$17.6 arcsec$^{2}$) around Q2112+059. The upper right
panel shows the same area after the final subtraction of the PSF. The
residual image in the upper right frame shows a fairly large
extended object (marked S), presumably a galaxy. The centroid of this
object coincides with that of the QSO to within 0.4 pixel (0.07
arcsec). The isophotes of the residual show no clear evidence for
ellipticity, rather they are consistent with $e=0$. There are two
additional objects that are both
consistent with being point sources 5 and 7 arcsec east of S, and
patches of faint extended fuzz both east and west of the object S.
In the lower panels of Fig.~\ref{fig3} we show for comparison the
field star C (0.35 mag fainter than Q2112+059) before and after 
PSF--subtraction.

\section{Spectroscopy from the HDA}
Our original aim with the observations was to detect the optical
counterpart of the DLA, but the perfect alignment of the galaxy S
with the QSO reported above suggests that we probably have detected the
QSO host galaxy rather than the galaxy causing the reported
candidate DLA absorption. To
help clarify this we searched the Hubble Data Archive (HDA) for
spectroscopic data of Q2112+059. In this section we describe the results
of this search.

\subsection{Retrieval of the spectra and Data Reductions}
We found two data sets suitable for this study. Goddard High Resolution
Spectrograph
(GHRS) spectra over the region from 1330 \AA \ to 1610 \AA \ (grism G140L)
was obtained in 1995 (PI Lanzetta) and in 1992 spectra of the region from
1610 \AA \ to 3250 \AA \ was obtained with the Faint Object Spectrograph
(FOS) (PI Bahcall).  Only the FOS-spectra has been published (Jannuzi et
al. 1998).

We retrieved the calibrated GHRS spectrum from the archive. Hence, the data
reduction was limited to extracting the 1 dimensional spectrum and
1$\sigma$ error spectra from the calibrated science data files and using the
calibrated wavelength solution files.

\subsection{Results}
    Fig.~\ref{fig4} shows the spectrum of Q2112+059 covering the region
from 1425 \AA \ to 1505 \AA. Also 
shown is the 1$\sigma$ error per 0.72\AA \ bin (lower curve).
\begin{figure}[t]
 \begin{center}
 \epsfig{file=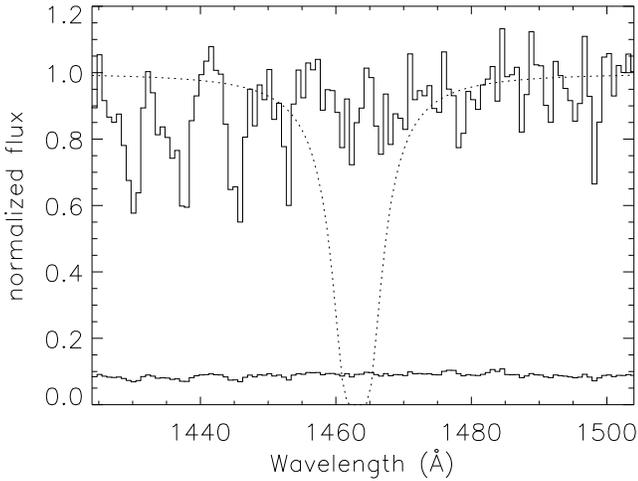,width=9cm}
 \end{center}
\caption{The 80\AA \ region around Ly$\alpha$ at z=0.2039 of the 
normalized and binned spectrum of Q2112+059. Also shown is the 
1$\sigma$ error per 0.72\AA \ bin (lower curve). The dotted line
shows a damped 
Ly$\alpha$ absorption line with a \ion{H}{I} column density of 
2$\times$10$^{20}$cm$^{-2}$ at a redshift of z=0.2039. The presence 
of such a line a excluded at the 29$\sigma$ level.}
\label{fig5}
\end{figure}

The presence of a damped Ly$\alpha$ absorption line in the Ly$\alpha$ 
forest of Q2112+059 at $z_{abs}=0.2039$ is excluded at the 29$\sigma$
confidence level. The FOS spectrum shows absorption due to
gas in the Milky Way and intervening metal line systems at redshifts 
z=0.370 and z=0.418 as well as metal line systems associated with 
Q2112+059, but no absorption lines at the redshift z=0.2039 
(Jannuzi private communication; Turnshek et al. in prep.).

As discussed by Rao \& Turnshek (2000) only 2 of 14 candidate DLA 
systems from the {\it IUE} survey of Lanzetta et al. (1995) have 
been confirmed. Q2112+059 adds one to the list of non-confirmed
candidates.

\section{Nature of the host galaxy}
\begin{figure}[t]
\begin{center}
\epsfig{file=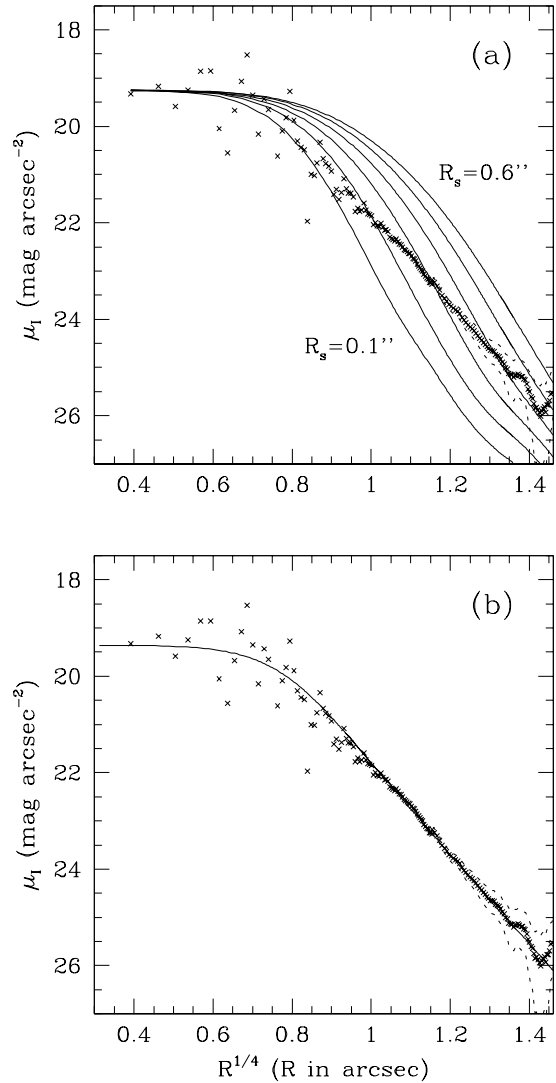,width=8.5cm}
\end{center}
\caption{{\it Upper panel:} The crosses show the surface brightness for 
the host galaxy
determined by azimuthal averaging on a 9 by 9 sub--pixel grid plotted 
against R$^{1/4}$ (in arcsec). Also plotted (full drawn lines) are 
6 exponential profiles with scale lengths 0.1--0.6 arcsec convolved with 
the PSF and azimuthally averaged in the same way as the data. 
As seen no exponential model provides a good fit to the data.
{\it Lower panel:}Same as the upper panel except here we have overplotted
the radial profile of a model consisting of an outer R$^{1/4}$ profile 
and a modified Hubble profile in the core (see text for details). As
seen this model match the data well.
}
\label{fig4}
\end{figure}

As there clearly is no DLA absorber in the sightline towards
Q2112+059, and as the centroid of the extended object we have
detected is closely aligned with the position of the QSO (to within
0.07 arcsec), we conclude that
we have detected the host galaxy of Q2112+059. We now return to
a discussion of what can be learned about the nature of this galaxy
from our deep imaging.

An attempt to fit a 2D galaxy model to the 2D PSF subtracted image shown
in Fig 3 was unsuccessful as the signal--to--noise of the image was
too poor to provide meaningful constraints. However, 
the alignment of the QSO with the centroid of its host, and the fact
that there is no evidence in the 2D residual image (Fig.~\ref{fig3})
for any elongation
of the extended object, suggest that we could significantly lower
the number of fitting parameters, and at the same time increase the
signal--to--noise of the data to fit, if we returned to the
1D regime. Again, as in the 2D case, we are left with
the degenerate problem of deciding how to scale the PSF we subtract
from the data. Galaxy cores are known often to have central spikes
(see e.g. M{\o}ller et al. 1995) which could be due either to central
point sources (AGN), to central density cusps or (which is the most
likely) to a combination of both
(for a discussion see Stiavelli et al. 1993). In the case of Q2112+059
there clearly is a central AGN (the QSO), which is the point source
we wish to separate out. Therefore, as in M{\o}ller et al. (1995) we
break the degeneracy by imposing the condition that the central profile
must be isothermal in the sense that the galaxy core is fitted by a
modified Hubble profile. This condition
closely corresponds to the requirement we used for the
2D PSF subtraction, namely that the residual must be ``flat'' or
``smooth'' in the centre.
With this assumption we effectively assign the maximum possible
flux to the central point--source, and therefore obtain a lower limit
to the flux of stellar origin. However, the central density cusps of
local early--type galaxies do not contain any significant fraction
of the total stellar light (1--10\% of the flux measured inside the
galaxy core, M{\o}ller et al. 1995), so the resulting uncertainty on
the flux from the total {\it stellar component} is insignificant.

For the 1D PSF we again used the weighted mean of the 5 field stars
(all field stars were given weights corresponding to their flux,
except star E which was given a lower weight because of the nearby
saturated star which caused variable sky across the profile).
In Fig.~\ref{fig4}a and Fig.~\ref{fig4}b we plot
the host surface brightness per square arcsec ($\mu_{I}$) against
R$^{1/4}$. Superimposed on the data are profiles corresponding to
a range of galaxy models. In Fig.~\ref{fig4}a we have superimposed 
six exponential disc profiles 
\[
I(R) = I_0 \times exp(-R/R_s)
\]
with scale lengths $R_s$ (0.1--0.6 arcsec), and in Fig.~\ref{fig4}b
the radial profile of a model with an R$^{1/4}$ outer profile 
\[
I(R) = I_0 \times \exp(-7.67 \times ((R/R_e)^{0.25}-1) ),
\]
and a modified Hubble profile 
\[
I(R) \propto (1+(R/R_c)^2)^{-1},
\]
in the core.
The models shown were produced on a 2D sub--pixel grid and convolved
with the observed 2D PSF. The final 1D model profiles were then
determined as azimuthal averages using the same procedure
as for the data.

It is seen that the radial profile of the host is noisy
in the central 0.5 arcsec where the errors from the PSF subtraction are
large and at radii larger than 3.5 arcsec where the uncertainty in the
measurement of the sky--level becomes dominant (sky--level errors are
shown as short dash curves in Fig.~\ref{fig4}). At radii between 0.5
and 3.5 arcsec the profile is very well determined.

Fig.~\ref{fig4}a shows that it is not possible to obtain an acceptable
fit with an exponential disc profile for any scale length. In contrast,
as seen in Fig.~\ref{fig4}b, a model with an R$^{1/4}$ outer profile with 
an effective radius of $R_e = 0.53$ arcsec (3.6 kpc) and a modified Hubble 
profile with a core radius of $R_c = 0.07$ arcsec (0.5 kpc) in the core 
provides an excellent fit.

\begin{table}[t]
\caption{Size, absolute magnitude and redshift of the host of
Q2112+059, and of Radio Galaxies (RG), Radio Load Quasars 
(RLQ) and Radio Quiet Quasars (RQQ) in the work by McLure et al.
(1999).}
\begin{center}
\begin{scriptsize}
\begin{tabular}{@{}lcccccc}
object    & $\bar{R}_e$ ,rms  & L$_{gal}$/L$_{QSO}$  & $\bar{M}_{gal}$ , rms & z \\
          & kpc   &                  &           &      &          \\
\hline
Q2112+059 & 3.6                 & 0.035                & $-$23.6          & 0.457     \\
RG        & 14.9  ,    5.4      & 25--3000             & $-$23.65  , 0.50 & 0.174--0.244\\ 
RLQ       & 11.2  ,    4.5      & 0.17--1.3            & $-$23.89  , 0.29 & 0.173--0.258\\
RQQ       &  9.8  ,    4.8      & 0.06--5              & $-$23.27  , 0.52 & 0.115--0.239\\
\hline
\end{tabular}
\end{scriptsize}
\label{profileparams}
\end{center}
\end{table}

\section{Summary and Discussion}

Based on an IUE spectrum of Q2112+059 Lanzetta et al. (1995) reported a
candidate damped Ly$\alpha$ line at z=0.2039. Subsequent HST spectroscopy
(Fig.~\ref{fig5}) has not confirmed this. This result is important in
its own right as DLAs are very rare at $z<1$. In the sample of DLAs
presented in Wolfe et al. (1995) there are 10 DLAs at z$<$1.0. Of the
10 only 4 are at z$<$0.5, and one of those was the candidate DLA towards
Q2112+059.

   Having ruled out the presence of an intervening DLA at z = 0.2039, 
we find that we are forced to interpret the elliptical galaxy we
detect under the PSF of Q2112+059 as the host galaxy of the QSO. 
The host galaxy has an I-band magnitude of 18.8 (total magnitude
determined from the model fit), and 
an effective radius of 3.6 kpc. I=18.8 corresponds 
to M$_I^{obs} = -23.6$ (no k-correction) in the assumed cosmology
(Sect. 1).

McLure et al. (1999) find that both radio galaxies (RGs), radio loud
QSOs (RLQs) and bright radio quiet QSOs (RQQs) preferentially are hosted
by massive, bright elliptical galaxies.
Our detection of a bright elliptical galaxy hosting the radio
quiet Q2112+059 is consistent with this result.
In Table~\ref{profileparams} we compare our results to those of
McLure et al. (1999) and find that
the host galaxy of Q2112+059 is in the bright end 
of the absolute magnitude distribution, but has a smaller effective
radius than any RQQ host galaxy in their sample.

Schneider et al. (1983) studied the brightest cluster galaxies in 83
Abell clusters (with redshifts in the same range as the McLure et al.
samples). They found a tight correlation between the core radii
and effective radii. The host galaxy of Q2112+059 falls nicely on
this correlation, but is smaller than any of the brightest cluster
galaxies of the Schneider et al. sample.

Compared to the work on local early type galaxies by Kormendy (1977),
the effective radius of the host galaxy of Q2112+059 falls in the
overlapping region between ``compact'' and ``normal'' early type
galaxies.

Finally, since the host galaxy we have detected is a spheroid we can
estimate the mass of the black hole using the correlations derived
by Magorrian et al. (1998) assuming that the correlation found at z=0
is also valid at z=0.457. We derive a black hole mass of about 
9$\times$10$^{9}$ M$_{\sun}$.

\section*{Acknowledgments}
We thank B. Januzzi for providing us with results on metal absorption
lines in the FOS spectrum of Q2112+0519 prior to their publication.
We are grateful to our referee Dr. Villar--Martin for suggestions and
comments which helped clarifying our manuscript on several important
points and significantly improved the presentation of our results.

\end{document}